\documentclass[10pt, conference, compsocconf]{IEEEtran}
\usepackage[caption=false]{subfig}

\usepackage{algorithm}
\usepackage{algorithmic}
\usepackage{dirtytalk}
\usepackage{color,soul}
\let\labelindent\relax
\usepackage[shortlabels]{enumitem}
\newcommand{\subparagraph}{}
\usepackage{titlesec}
\usepackage{xcolor}

\definecolor{darkgreen}{rgb}{0.2,0.4,0}

\makeatletter
\renewcommand{\fnum@figure}{Fig.\thefigure}
\makeatother

\ifCLASSINFOpdf
   \usepackage[pdftex]{graphicx}
\else
\fi

\begin{document}
\title{\vspace*{-1cm}Swarm-based Drone-as-a-Service (SDaaS) for Delivery\vspace*{-0.5cm}}





%

\author{\IEEEauthorblockN{Balsam Alkouz\IEEEauthorrefmark{1},
Athman Bouguettaya\IEEEauthorrefmark{1},
and Sajib Mistry\IEEEauthorrefmark{2}
}
\IEEEauthorblockA{\IEEEauthorrefmark{1}School of Computer Science\\
The University of Sydney, Australia\\ 
\{balsam.alkouz, athman.bouguettaya\}@sydney.edu.au}
\IEEEauthorblockA{\IEEEauthorrefmark{2}School of Electrical Engineering, Computing and Mathematical Sciences\\
Curtin University, Australia\\
sajib.mistry@curtin.edu.au}\vspace*{-0.5cm}}



\maketitle

\begin{abstract}
We propose a novel framework for composing Swarm-based Drone-as-a-Service (SDaaS) for delivery. Two composition approaches, i.e., sequential and parallel are designed considering the different behaviors of drone swarms. The proposed framework considers various constraints, e.g., recharging time and limited battery to meet delivery deadlines. We propose SDaaS composition algorithms using a modified A* algorithm. A cooperative behavior model is incorporated to reduce recharging and waiting time in a delivery. Experimental results prove the efficiency of the proposed approach.

\end{abstract}
\vspace{-0.1cm}

\vspace{-0.1cm}

%
\IEEEpeerreviewmaketitle

\section{Introduction}
\vspace{-0.1cm}
Drones are aircrafts without a human pilot that operate with various degrees of autonomy \cite{circular2011328}. The wide availability of drones opens opportunities for a large number of applications including disaster management, crowd control, and agriculture \cite{chmaj2015distributed}. These abundant applications come as a result of the drop in prices and increased sophistication of drones. Drones are commonly used for sensing, inspection, and delivery of packages \cite{shahzaad2019composing}. Our \emph{focus} is on the use of drones in delivering goods typically within a city limit.

There has recently been an increasing interest in drone delivery services in the industry \cite{bamburry2015drones}. For example, Amazon utilizes quadcopters to fly packages to customers within 30 minutes\footnote{Amazon Prime Air. https://www.amazon.com/Amazon-Prime-Air/}. Corporate based merchants now use Googles project Wing to deliver goods in Canberra\footnote{Google Wing - Canberra. https://wing.com/australia/canberra/}. The use of drone has the added value of being more \emph{convenient} and \emph{environmentally friendly} as it delivers faster and uses less energy.\looseness=-1

Drone delivery services exhibit the same behavior with the non-functional properties of the service paradigm. Hence, this paradigm is the natural fit for \emph{Drone-as-a-Service} \cite{shahzaad2019composing}. In this respect, each drone delivery service has functional and non-functional (QoS) properties. The \emph{functional} aspect of a drone delivery service is expressed as the delivery of packages from a source to a destination using drones. The \emph{non-functional} properties represent the drones' energy consumption, delivery cost, delivery time, etc. A model was proposed for Drone-as-a-Service where the authors abstract a drone travelling in a line segment in a skyway network as the service \cite{shahzaad2019composing}.

There are  several instances where a \textit{swarm} of drones may be needed to fulfill the requirements of a delivery. A scenario, where a swarm may be needed, is when there is a need to deliver goods but a single drone cannot carry the package. This may be because of the weight of the package that \emph{exceeds a drone’s payload capacity}. For instance, a customer may request a heavy electronic gadget that is made up of separable lighter pieces. Therefore, we need multiple drones to deliver the pieces. A swarm additionally would be needed when \emph{multiple different items} are asked to be delivered at a certain time and location. For example, a hospital may need multiple different medical pieces of equipment delivered as fast as possible in cases of emergencies \cite{scott2017drone}. In this case, multiple drones are needed to deliver the pieces of equipment. An additional scenario is when multiple single drones are available, and a single drone can carry the package weight, but the energy consumption due to the payload will not be enough to cover the distance. Here, \emph{multiple drones could be used to ensure the overall energy burden is shared} due to the payload sharing \cite{akram2017security}. 

While the drones provide multitude of opportunities, there are some intrinsic and extrinsic challenges in drone delivery. These challenges are mostly related to limited battery lifetime, payload, endurance, flight range, weather, obstacles, security, and public safety. Additionally, there are added challenges that face swarm-based delivery that include coordination and distribution \cite{campion2018uav}. Hence, a swarm-based service composition model is needed to address these challenges. A swarm is a set of drones that move together limited by a time and space window \cite{cimino2015combining}. A swarm moves together from a source to a destination in a delivery.\looseness=-1 

We define a \emph{Swarm-based Drone-as-a-Service (SDaaS)} by its \emph{functional} and \emph{non-functional (QoS) properties.} The functional property represents the delivery of multiple packages by a swarm between source and destination nodes in a skyway network. A skyway network is made up of sky segments connecting nodes in the connected network \cite{shahzaad2019composing}. Examples of non-functional properties include delivery time, payload, energy consumption, etc. Given a skyway network, \emph{our aim is to compose the best set of skyways for delivery}. The best set is the set that optimizes the QoS properties like energy consumption, and delivery time. We focus on optimizing the \emph{delivery time}.\looseness=-1  

\begin{figure}[!t]
\centering
\includegraphics[width=3.5in]{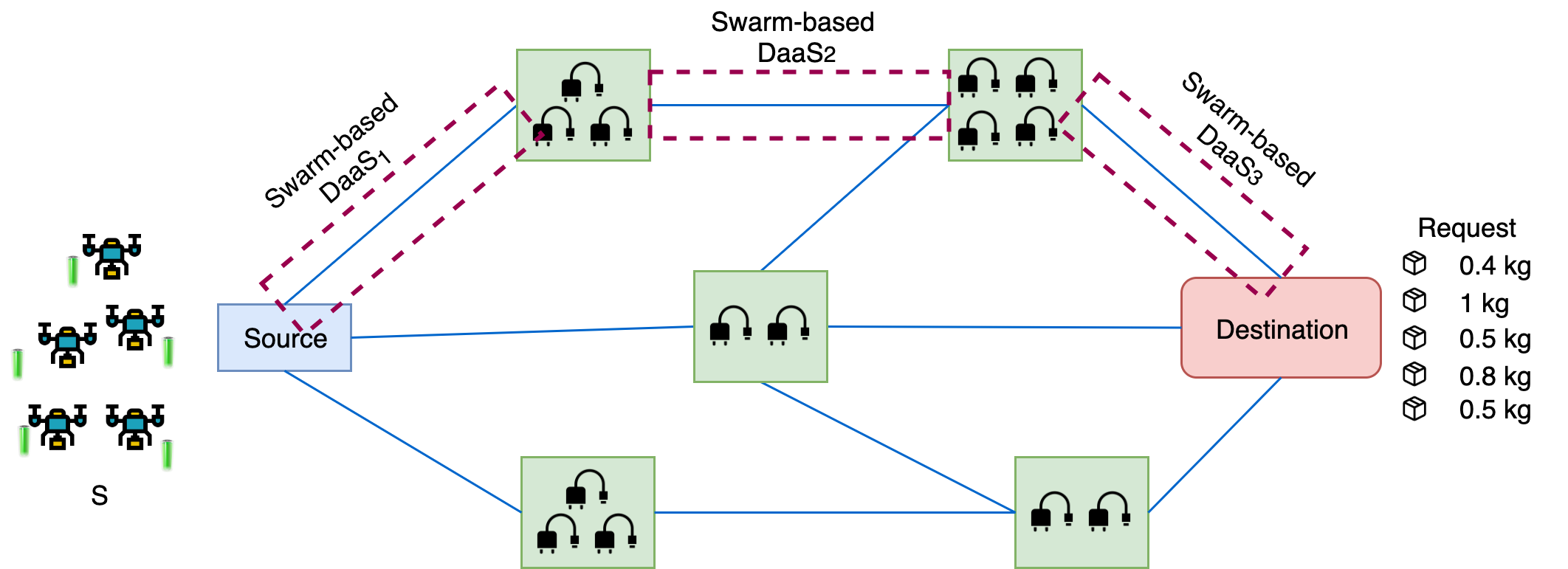}
\caption{Skyway network for swarm-based drone delivery}
\label{fig_skyway}
\end{figure}

We consider a \emph{constrained swarm-based drone delivery as a motivating scenario}. \label{scenario} A hospital requires medical supplies in large quantities daily. However, a single drone cannot carry all the supplies at once due to its payload limitations. Delivering medical supplies to hospitals is a very time sensitive problem. Furthermore, the supplies need to arrive as fast as possible in cases of emergencies. Some medicines need other compatible medicines to work with, e.g. Synergistic Drugs. Hence, a swarm of drones is needed to deliver all the  medicines at the same time. Fig \ref{fig_skyway} presents the skyway network between the supplier and the destination point. We assume that the intermediate nodes have different numbers of recharging pads. We also assume that the packages requested are of different weights. The weight of the package is directly proportional to the drones energy consumption \cite{d2014guest}. In addition, we assume that the maximum weight of the package is under the drones payload capacity.\label{maxweight} This in return will affect the distance a drone can travel. \emph{Given all these constraints, the swarm must deliver the packages from the source to the destination as fast as possible.} As mentioned earlier, the packages should arrive at approximately the same time. Hence, the swarm should also comply to a limited time window between the arriving packages at the destination. We assume that the environment is deterministic, i.e., we know in advance about
the packages weights, drones speeds, battery capacities, and battery consumption rates. \label{failures}Future work will consider uncertainties in the environment like weather conditions and damages. Damages could include broken charging stations and mechanical problems in the drones which may lead to failures.\looseness=-1 


\setlength{\textfloatsep}{0pt}

We summarize the key contributions of this research as follows:
\begin{itemize}[leftmargin=*]
    \item A novel Swarm-based Drone-as-a-Service (SDaaS) model.
    \item A categorization of different types of SDaaS compositions: Sequential and Parallel. These types are derived from different behaviors a swarm can exhibit.
    \item Constraint-aware SDaaS using a modified A* heuristic algorithm to find the optimal services compositions for every type.
    \item A cooperative enhancement to the SDaaS composition. Here, the drones in the swarm behave in a way that improves the ultimate goal of fast delivery.
\end{itemize}


\vspace{-0.2cm}
\section{Related Work}
\label{relatedwork}
\vspace{-0.1cm}

Several studies used drone swarms in different applications. The most common applications include target search, entertainment, and building airborne communication networks. The targets in a target search application may be toxic clouds \cite{avvenuti2018detection}, parasites \cite{potrino2019drones}, or  humans \cite{rivera2016post}.  Generic drone target search approaches are also discussed in the literature \cite{cimino2015combining} . Another application of drones explored in the literature is building communication networks. These networks may be useful to connect with rural areas or disaster damaged areas \cite{shi2018drone}. \looseness=-1 

Multiple studies addressed the challenges in single drone delivery. An article highlighted  the favorable and unfavorable factors in drone delivery \cite{bamburry2015drones}. The article discussed with examples the benefits of drone delivery when speed is critical. For instance, harvested crops may be delivered from the farm to the warehouse faster. Another example presented is the delivery of medical equipment to remote and hard-to-reach areas. A feasibility analysis was carried out on adopting drones for delivery \cite{d2014guest}. The author studied the feasibility in terms of power and energy consumption, and cost efficiency. He concluded that the cost of a delivery flight, including the battery cost and life-time, is much lower than the cost of traditional delivery. A new model for Drone-as-a-Service for delivery purposes, using a single drone, was proposed \cite{shahzaad2019composing}. Their objective was to select and compose the best drone services for delivery i.e. minimize the delivery time, from a set of requests. The authors continued to address the challenges in DaaS composition with the focus on the constraints in the intermediate nodes \cite{shahzaad201constraint}.\looseness=-1 

Several papers discussed how a swarm should be \emph{coordinated} in different applications. All swarms of drones exhibit similar features and may be defined by several \emph{aspects}. Swarms of drones may be classified into different types based on the swarm members \emph{structure}. Three main types are recognized in the literature namely: Static, Dynamic and  Hybrid \cite{akram2017security}. In terms of \emph{movement}, there are several ways introduced in the literature about how drones in a swarm can move while staying close \cite{avvenuti2018detection} . The first is a random movement while maintaining the distance between individual drones.  The second is stigmergy, which is inspired by ants that leave pheromones to attract other ants. In the same manner, drones are attracted to digital pheromones that could be released by a master drone. The third is the flocking movement, which is based on the rules of alignment, separation, and cohesion of a bird’s flock. A swarm of drones is expected to have some level of autonomy. 

 
 
For delivery, the work done in swarm-based drone delivery is still in its infancy. In addition, the papers that addressed drone swarm deliveries define a swarm as multiple single drones that deliver different requests \cite{san2016delivery}. We define a swarm as a set of drones that act as a single entity. An approach was proposed to assign a swarm of UAVs to deliver items with regards to some constraints \cite{san2016delivery}. A swarm is made up of multiple single drones managed to deliver from one source to multiple destinations. The authors adopt Genetic Algorithm to optimize the selection of drones for delivery.  The service paradigm is leveraged to model other types of travel services like transport \cite{neiat2017crowdsourced}. To the best of our knowledge, there is no previous work done on utilizing a swarm of drones for delivery. \emph{This paper is hence, the first attempt to model swarm-based drone delivery services in a highly constrained environment.}
 

\section{Swarm-based Drone-as-a-Service Model}
\label{sdaas}

\begin{figure}[!t]
\centering
\includegraphics[width=\linewidth]{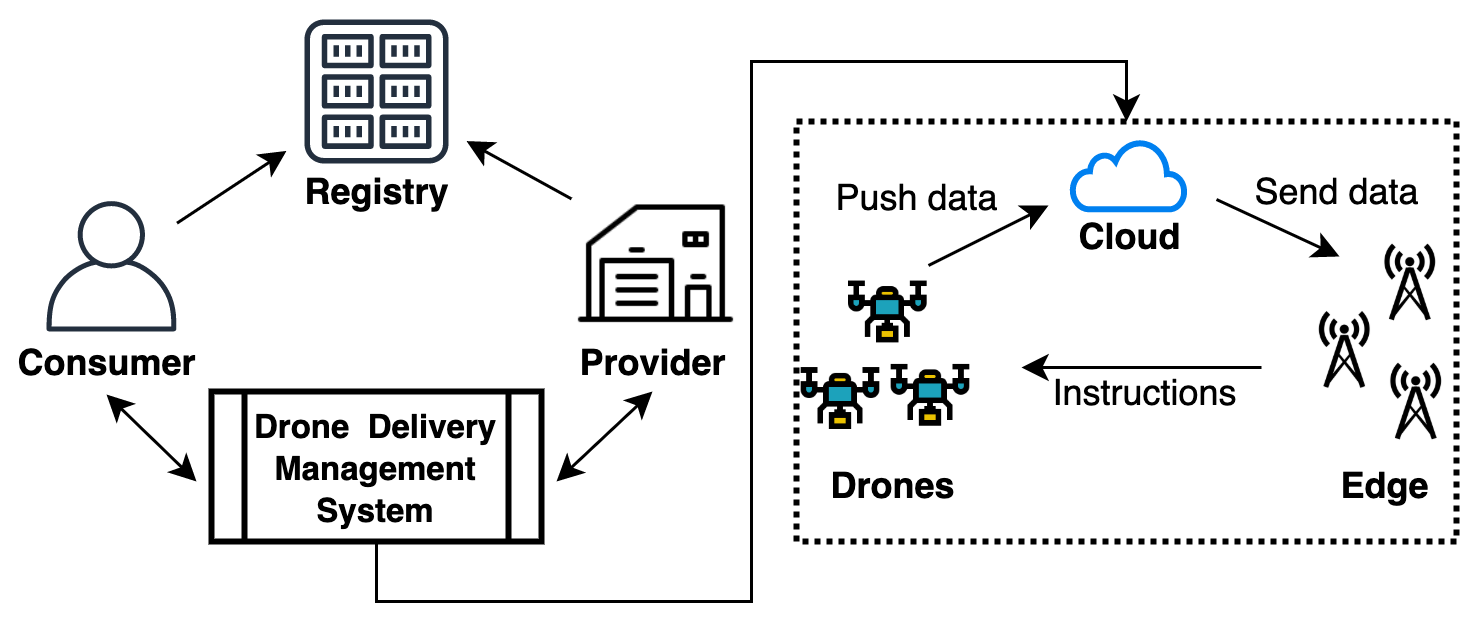}
\caption{Swarm-based DaaS system architecture}
\label{architectire}
\end{figure}

We present the basic system architecture for swarm-based delivery services.\label{architecture} The architecture is premised on having a set of drones that deliver goods from a source to a single destination. The architecture as shown in Fig.\ref{architectire}  is an adapted service oriented architecture (SOA) with a drone delivery management system. The \emph{provider} publishes its services to the \emph{registry} to advertise them. The \emph{consumer} locates the services in the \emph{registry} and invokes them. When the consumer invokes a service, the drone delivery management system determines the set of drones needed and the path to be taken. We assume that the computations will be shared by \emph{drones, edge nodes, and cloud}. Simple calculations will be performed at the \emph{drone} level. More evolved computations will be done at the \emph{edge} level. The edge nodes are distributed in strategic locations to aid delivery. Computations requiring lots of data will be done on the \emph{cloud}. As the drones move, they communicate their locations and battery states to the cloud as it holds all the data about the swarm and the request. The system at the edge nodes makes quick computations when needed. The edge nodes continuously take the swarm location and battery data from  the cloud. The decisions made by the system at the edge nodes are communicated back to the drones to instruct them on where to go. The drone function is a dichotomy of telling the cloud where it is and receiving from the edge where to go. Once the swarm reaches the destination, a confirmation message is sent to the consumer to pick the packages. Consumers living in apartment buildings may have their packages dropped at one landing spot and a concierge service from the building manages these packages \label{concierge} \footnote{https://www.unmannedairspace.info/urban-air-mobility/first-london-apartment-block-drone-delivery-landing-site-open-june-2019/}  or the landing spot may be in the customer's balcony \cite{brunner2019urban}.


\emph{We abstract each swarm travelling on a skyway between two nodes as an SDaaS} (see Fig. \ref{fig_skyway}). Drones have limited battery capacities and flight ranges \cite{d2014guest}. Therefore, a swarm may require recharging at intermediate nodes for long distance deliveries. Our goal is to \textit{select} and \textit{compose} the best set of swarm-based drone services from a source to a destination with an optimal delivery time. We should also consider the highly constrained environment around the swarm-based delivery including the limited time window between the packages arrival. In this paper, we consider the environment to be deterministic. 

We formally define a Swarm-based Drone-as-a-a-Service (SDaaS). Then we define an SDaaS customer request.\\
\textbf{Definition 1: Swarm-based Drone-as-a-a-Service (SDaaS).} An SDaaS is defined as a tuple of $<SDaaS\_id, S, F>$, where
\begin{itemize}
    \item $SDaaS\_id$ is a unique identifier
    \item $S$ is the swarm/subswarm travelling in SDaaS. $S$ consists of $D$ which is the set of drones forming $S$, a tuple of $D$ is presented as $<d_1,d_2,..,d_n>$. $S$ also contains the properties including battery levels of every $d$ in $D$ $<b_1,b_2, ..,b_n>$, the payloads every $d$ in $D$ is carrying $<p_1,p_2,..,p_n>$, and the current node $N$ the swarm S is at.
    \item F describes the delivery function of a swarm on a skyway segment between two nodes, A and B. F consists of the travel time $tt$,  charging time $ct$, and waiting time $wt$ when recharging pads are not enough to serve $D$ simultaneously in node B.
\end{itemize}

\textbf{Definition 2: SDaaS Request.} A request is a tuple of $<\alpha, \theta, P>$. $\alpha$ is the source node, $\beta$ is the destination node, and $P$ are the weights of the packages requested, where $P$ is $<p_1,p_2,..p_n>$.\\

\vspace{-0.5cm}
\subsection{Types of behaviors}
\label{behaviors}
We identify two different types of behaviors of swarm-based drone services composition. The two types are derived from the different types of swarms: static and dynamic swarms \cite{akram2017security}. These behaviors map to two different types of service compositions: \emph{Sequential and Parallel}.

\begin{figure*}[ht]
\centering

\subfloat[Static swarm with sequential composition]{\includegraphics[width=0.46\linewidth]{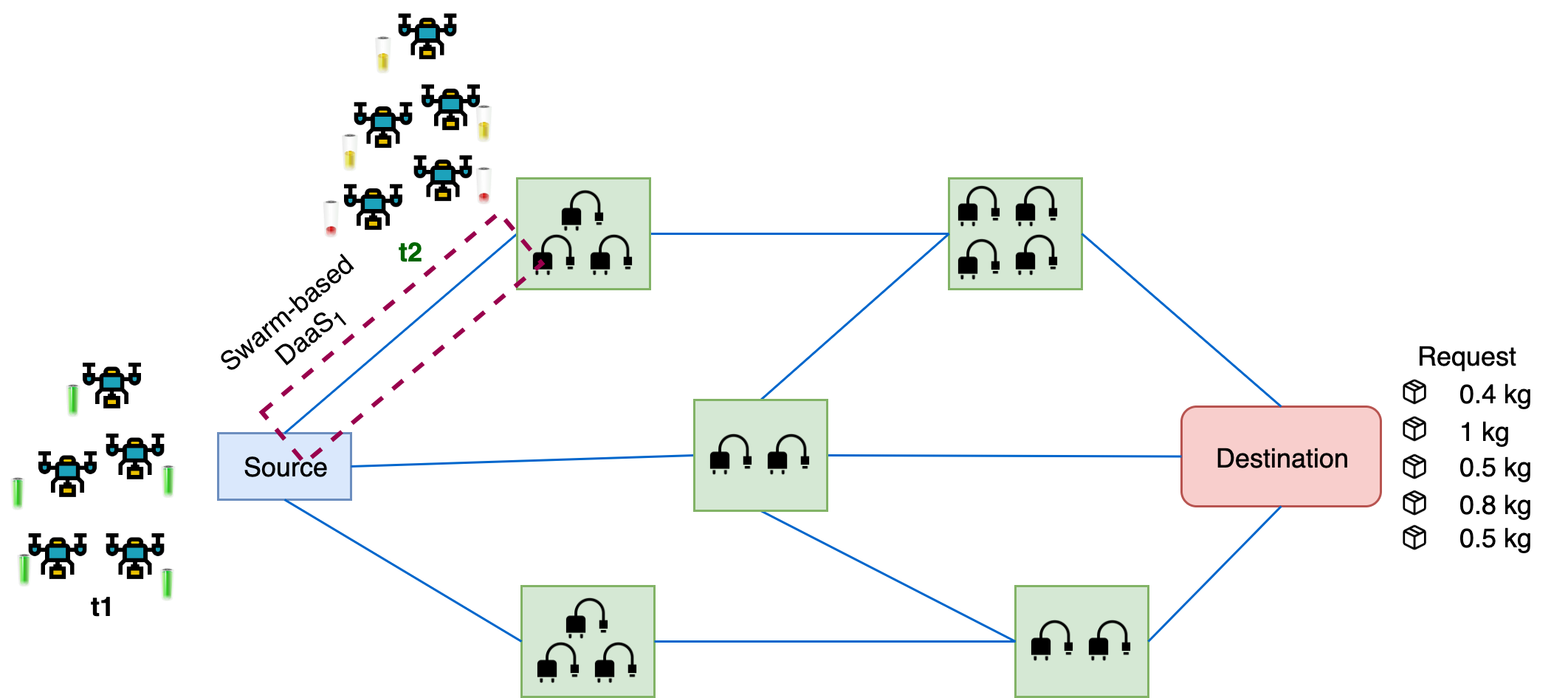}\label{fig_static}\setlength{\belowcaptionskip}{-5pt}
}\quad
\subfloat[Dynamic swarm with parallel composition]{\includegraphics[width=0.46\linewidth]{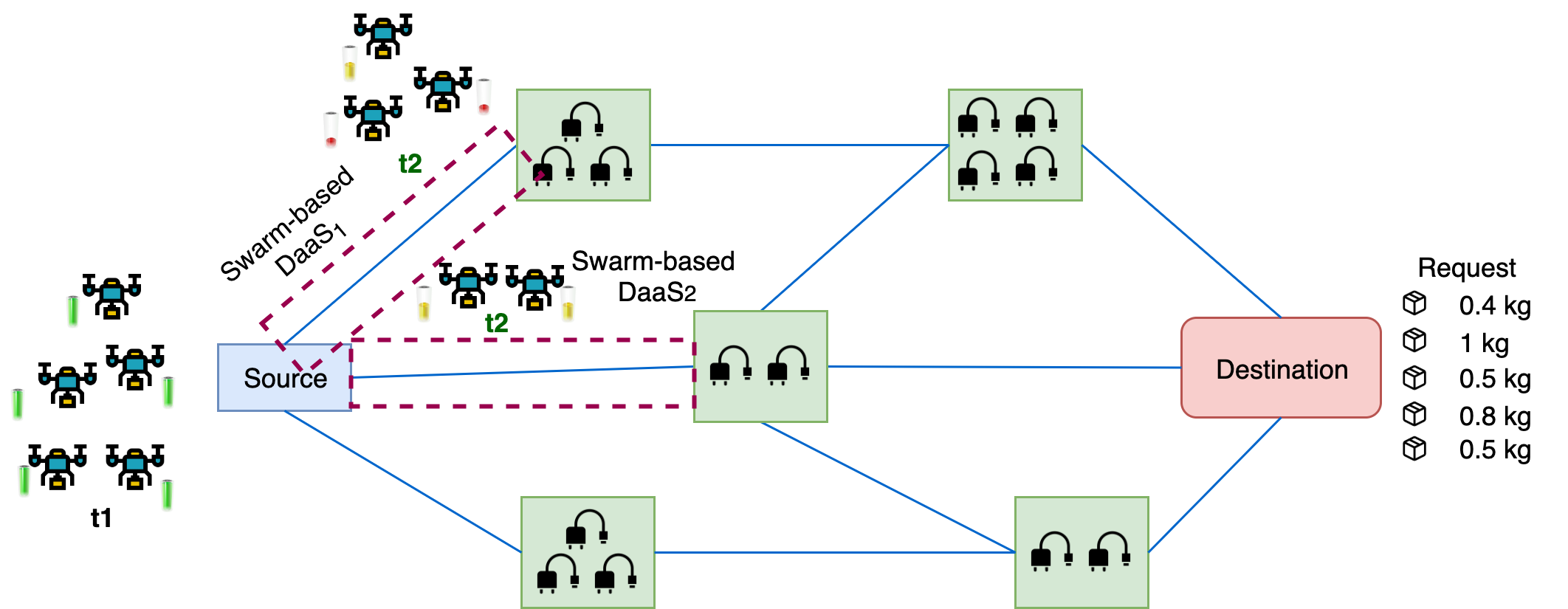}\label{fig_dynamic}\setlength{\belowcaptionskip}{-5pt}
}
\setlength{\belowcaptionskip}{-10pt}
\caption{Static and Dynamic swarm-based services snapshots at different times}
\label{fig:test}
\vspace{-3mm}
\end{figure*}

\begin{enumerate}[leftmargin=*]
    \item \emph{Static swarms:} A static swarm is a swarm of drones whose members are decided at the source \cite{akram2017security}. Therefore, no new enrollments or retirements of drones occur at intermediate nodes. The same set of drones form a swarm at the source node and traverse the network together till they reach the destination. This behavior ensures that all drones arrive to the destination at the same time. However, in the case of a highly constrained environment, the charging pads in the nodes may not be able to cater the needs of all the drones at the same time. This may cause delays and increased waiting times due to sequential recharges. In addition it may cause congestion at a node preventing other services from utilizing the path. We refer to this type of composition as a \emph{sequential swarm-based drone services composition}.  Fig \ref{fig_static} shows snapshots at different times for a static swarm-based drone delivery and the sequential service composition.
    
    \item \emph{Dynamic swarms:} \label{dynamic} A dynamic swarm is originally defined as a swarm of drones where the inclusion of new members as well as the leave of existing members is allowed \cite{akram2017security}. In our case, the swarm members are formed at the source, and when it traverses the network its original structure may change. However, the original members of the swarm are still the same. In case of delivery, swarms at different nodes may split to form sub-swarms or merge together. We call this process \emph{banding and disbanding} of a swarm. Adopting this type ensures that the resources in the network are distributed amongst the drones. As a result, parallel charging occurs which reduces the waiting time and congestion. When adopting this method, we have to ensure that all the sub-swarms will meet at the destination within a limited time window.  A sub-swarm is made up of at least two drones and hence is considered a swarm. Having multiple sub-swarms at the same time means that we will have \emph{parallel swarm-based drone services compositions}. Fig \ref{fig_dynamic} shows snapshots at different times for a dynamic swarm-based drone delivery and the parallel service composition.
    
\end{enumerate}

We experiment the two aforementioned behaviors on different requests. We aim to evaluate the efficiency, in terms of delivery time, of adopting static and dynamic behaviors. We identify the following constraints on the swarm-based drone delivery services:

\begin{itemize}[leftmargin=*]
    \item \emph{Limited arrival time window}: The ultimate goal is to deliver all the packages at approximately the same time, as fast as possible. Therefore, the arrival is limited by a time window $w$, which is the time between the first arriving package $p_1$ and the last arriving package $p_n$. $w=p_1-p_n$
    \item \emph{Different recharging requirements between drones}: The weight of the packages in a single request are different. Therefore, every drone carries a different weight. This difference in payloads results in different energy consumption rates for every drone in a single swarm. Hence, decisions on swarm-based drone services selection should cater the needs of all drones in a swarm.
    \item \emph{Limited charging capacities at nodes}: A node in a skyway network could be the source, the destination, or an intermediate charging station. All drones are charged fully at the source node. However, to deliver for long distances the drones need to recharge at the intermediate charging stations. We assume that the process is automated using wireless recharging stations, i.e. no human interaction is possible to replace drones batteries.\label{replacebattery} Every charging station has a different number of charging pads. Consequently, an intermediate node should be selected to cater the needs of all drones in a swarm.
\end{itemize}

Our focus is on the composition of swarm-based drones services considering the aforementioned constraints to ensure a fast delivery. To the best of our knowledge, existing approaches focus only on single drone deliveries. None of the existing approaches focus on swarm-based drone deliveries.\looseness=-1
\vspace{-0.2cm}
\section{SDaaS Composition Framework}
\label{framework}
\vspace{-0.1cm}

There are two types of compositions that we tackle in this paper: Sequential and Parallel. We create two heuristic algorithms to tackle each composition type. The heuristic algorithms are inspired by the A* algorithm \cite{hart1968formal}. We don't only consider the cost to reach the neighboring node, i.e. the distance. At every node, when selecting the next best neighboring node we additionally consider the heuristic value of the node. In our case, the heuristic value is the charging time and the waiting time at the node. 

\subsection{Sequential SDaaS Composition}
In a \emph{sequential composition} all the drones $D$ form a swarm at the source node and traverses the network to the destination while staying together. The number of drones $D$ in $S$ is equal to the number of packages $P$ in a request $R$. While the swarm is not at the destination node the swarm computes the potential to reach the destination from its current node without stopping to recharge. We compute the potential based on the payload all drones are carrying which affects the battery level and the total distance a drone can travel. If the destination is reachable then the swarm traverses the nodes till it reaches the destination and the travel time is updated. If the destination is unreachable then the swarm finds the nearest reachable neighbor, with lookahead $l$, from the current node with the minimum travel time and node time. The node time constitutes of the charging time to 100\% and the waiting time of drones waiting to get charged sequentially. $NT=ct+wt$. As the payload affects the battery consumption, the charging time between the different drones in the swarm is different. We take the maximum charging time $ct$ of the drones to represent the charging time of the swarm.  The swarm then again tries to find if the destination is reachable directly until the swarm is at the destination node. The total delivery time is the total travel time and node time. Algorithm \ref{sequntialAlg} describes the sequential service composition process.\looseness=-1

\begin{algorithm}
 \caption{Sequential Services Composition Algorithm}
 \label{sequntialAlg}
 \small
 \begin{algorithmic}[1]
 \renewcommand{\algorithmicrequire}{\textbf{Input:}}
 \renewcommand{\algorithmicensure}{\textbf{Output:}}
 \REQUIRE $S$, $R$
 \ENSURE  $t$
 \STATE $t$ = 0
    \WHILE{$S$ is not at destination}
        \STATE distance to destination= \textbf{Dijkstra}(current, destination)
        \STATE \textbf{compute} energy consumption for every $d$ in $S$ based on $R$ package weights and distance to destination
        \IF{all $d$ in $S$ can reach destination without intermediate nodes}
        \STATE $S$ travels to destination
        \STATE $t$+=travel time
        \ELSE
        \STATE \textbf{find} nearest neighbor nodes where $lookahead =l$
        \STATE \textbf{select} best neighboring node (min travel time and min charging time)
        \STATE $S$ travels to neighboring node
        \STATE $t$+=travel time + charging time + waiting time
        \ENDIF
    \ENDWHILE
 \RETURN $t$
 \end{algorithmic}
 \end{algorithm}

\subsection{Parallel SDaaS Composition} 
\label{parallelcompo}
In a \emph{parallel composition}, there can be multiple sub-swarms in the network at the same time as the initial swarm may disband into $n$ sub-swarms. Every sub-swarm must consist of a minimum of two drones. \label{minsize} We limit the minimum numbers of drones in a swarm but not the maximum. \emph{We treat every sub-swarm as a full swarm.} For every swarm $<s_1,s_2,..,s_n>$  in the skyway network that didn't reach the destination three scenarios may happen. Algorithm \ref{algparallel} presents the different scenarios explained below.

\begin{enumerate}[leftmargin=*]

    \item \emph{Scenario 1: All drones can reach the destination without recharging at intermediate nodes.}
    The measurement of the potential in reaching the destination is similar to the method used in the sequential composition. In this scenario, the swarm travels together traversing the nodes till it reaches the destination. The travel time is added to the total travel time.\looseness=-1
    \item \emph{Scenario 2: A sub-swarm can reach the destination without recharging at intermediate nodes.}
    In this scenario, the swarm divides into two sub-swarms, one that goes directly to the destination in a similar manner to scenario 1 and the second stays at the current node, and gets treated as scenario 3. In this scenario, we should ensure that the two sub-swarms are of a minimum size two , i.e. $D>2$. The travel time of sub-swarm one is added to the total delivery time. \looseness=-1
    \item \emph{Scenario 3: No sub-swarm can reach the destination without recharging at intermediate nodes.}
    In this scenario, we split the swarm S into all possible combinations of $<s_1,s_2,..s_n>$ where the maximum number of $max\_splits = x$. If $x=2$, a swarm may disband to a maximum of two sub-swarms at a node. We also get all the neighboring nodes of the current node where $lookahead = l$. Lookahead is the level of neighboring nodes from the current node. If $l = 1$, then the directly connected nodes and the 2\textsuperscript{nd} level connected nodes are retrieved. Then, we select the best split that adds the minimum time to the total travel time, i.e., minimum travel time, and minimum charging and waiting times. Every combination consist of a maximum $x$ sub-swarms that can go to a maximum of $x$ neighboring nodes. Once the best combination is selected, the sub-swarms travel to the selected neighboring nodes and charge to 100\%. \label{limitedwindow}As described earlier, the arrival of the packages at the destination is limited by a time window $w$. If the time at the destination between the first arriving and the last arriving sub-swarm is less than $w$, then the sub-swarms that arrived first must wait for the rest of the swarm \emph{at the destination} adding to the waiting time. Otherwise, the first arriving sub-swarms wait \emph{at previous nodes} before travelling to the destination. The travel time, the charging time, and the waiting time (if exists) will be added to the total delivery time.\looseness=-1
\end{enumerate}

\begin{algorithm}
 \caption{Parallel Services Composition Algorithm}
 \label{algparallel}
 \small
 \begin{algorithmic}[1]
 \renewcommand{\algorithmicrequire}{\textbf{Input:}}
 \renewcommand{\algorithmicensure}{\textbf{Output:}}
 \REQUIRE $S$, $R$, $n$
 \ENSURE  $t$
 \STATE $t$ = 0
  \FOR {every $s$ in $S$}
    \WHILE{$s$ is not at destination}
        \STATE distance to destination= \textbf{Dijkstra}(current, destination)
        \STATE \textbf{compute} energy consumption for every $d$ in $s$ based on $R$ package weights and distance to destination
        \IF{all $d$ in $s$ can reach destination without intermediate nodes}
        \STATE $s$ travels to destination
        \STATE $t$+=travel time
        \ELSIF{some $d$ in $s$ can reach destination without intermediate nodes}
            \IF{$d >= 2$ \AND $s-d >=2$}
            \STATE $d$ travels to destination
            \STATE \textbf{charge} $s-d$ to 100\%
            \STATE $t$+=travel time + charging time + waiting time
            \STATE \textbf{update} $S$ (remove $s$ and add $s-d$)
            \ENDIF
        \ELSE
            \STATE split $s$ into  all combinations of $s1$..$sn$, where $d$ in ($s1$..$sn$) $>=2$ \AND $max\_splits = x $
            \STATE \textbf{find} nearest neighbor nodes where $lookahead =l$
            \STATE \textbf{select} best combination of swarm split and target nodes (min travel time and min charging time)
            \STATE $s1$..$sn$ travel to neighboring nodes
            \STATE \textbf{charge} $s1$ .. $sn$ to 100\%
            \STATE $t$+=travel time$_{s1 .. sn}$ + charging time$_{s1 .. sn}$ + waiting time$_{s1 .. sn}$
            \STATE \textbf{update} $S$ (remove $s$ and add $s1$ to $sn$)
        \ENDIF
    \ENDWHILE
  \ENDFOR
 \RETURN $t$
 \end{algorithmic}
 \end{algorithm}
We adopt the Boids algorithm to ensure that all the drones in a swarm travel together and are close to each other \cite{reynolds2006boids}. This algorithm is based on the flocking behavior of birds including alignment rules to ensure the drones are moving in the same direction. With separation rules, the drones keep a minimum distance between each other to avoid collisions. Finally with the cohesion rules, the drones tend to move towards the center of the swarm ensuring that they stay close together. We also fix the speed of the drones regardless of the payload they are carrying. \emph{We focus on the effect of payload on the energy consumption} rather than the  effect of speed.

\subsection{The Effect of Cooperation}
\begin{figure}[t]
\centering
\includegraphics[width=0.9\linewidth]{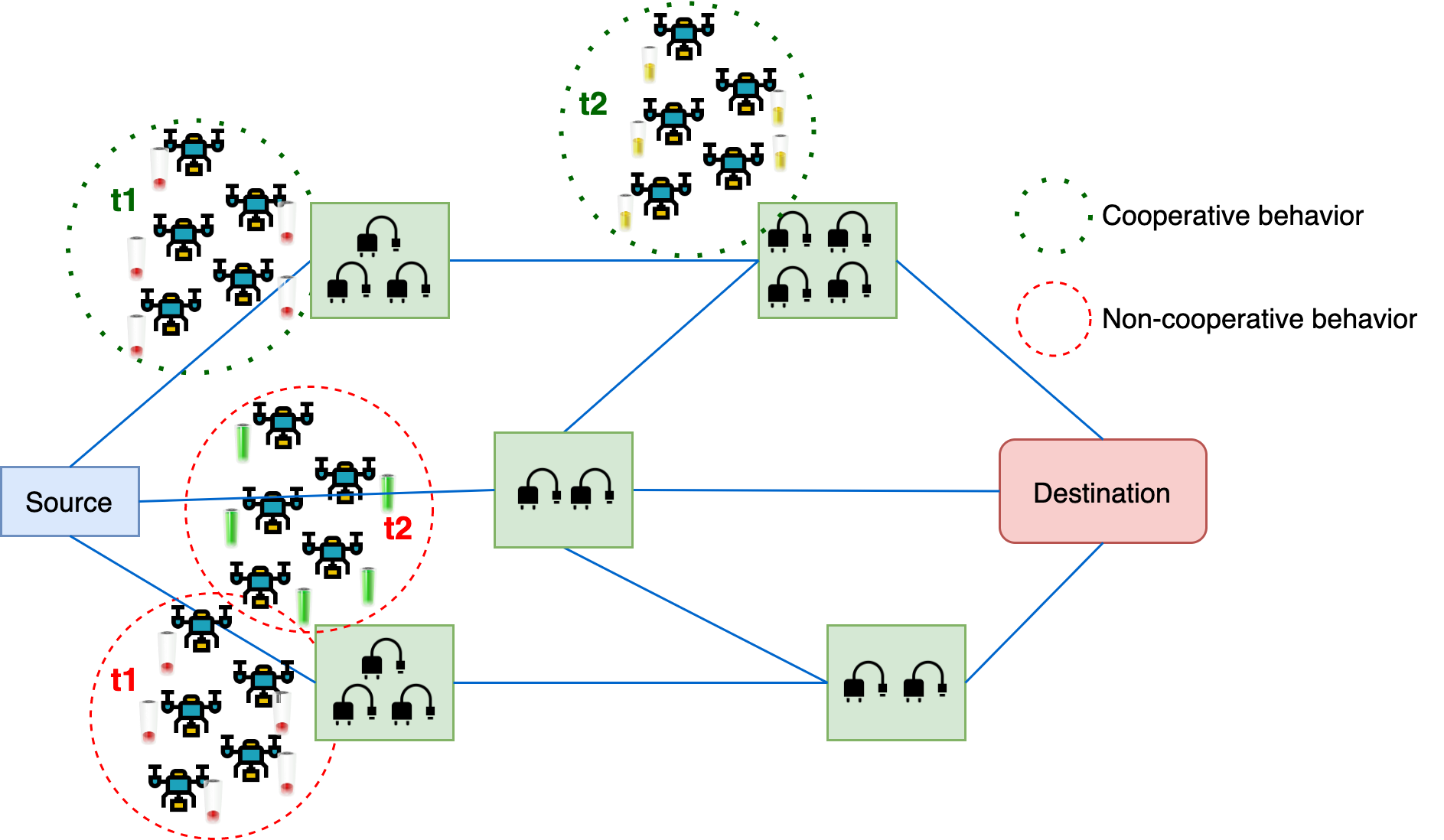}
\caption{The effect of cooperative behavior at different $t$}
\label{altruism}
\end{figure}

Cooperation in biology is the behaviour where groups of organisms work together to achieve a common benefit \cite{inbook}. This behavior was not only adopted in biology but also in economics which refers to situations in which the individuals seek win-win outcomes from working together \cite{fehr1999theory}. In the case of a Swarm-based Drone-as-a-service composition, cases could occur where multiple drones are sharing a limited number of recharging pads $rp$ at a node where $D>rp$. In this case, we study the effect of charging the drones in the swarm up to what takes them to neighboring nodes instead of having every drone charging to 100\%. In the case of a cooperative behavior, the drones at their own expense do not charge fully to achieve the common goal of fast delivery. Instead, the drones cooperate with the goal of reducing the charging time and waiting time at a node. We study if this cooperative behavior reduces the waiting time of sequential charging and ensures better distribution of resources. Fig. \ref{altruism} shows the difference between cooperative and non-cooperative behaviors at different timestamps. Because of cooperation, the waiting time is reduced and the swarm travels to the next node before the non-cooperative swarm.

\vspace{-0.2cm}
\section{Experimental Results and Discussion}
\label{results}
\vspace{-0.2cm}

In this section, we evaluate the performance of the proposed types of compositions against Dijkstra's algorithm and Brute Force approach. We conduct a set of experiments to evaluate the controlling attributes in the composition like the lookaheads of the retrieved neighboring nodes and the maximum number of disbands at a node. We also show the results before and after adopting a cooperative behavior.

\label{seqsdaascompo}
The dataset used in the experiments is an urban road network dataset from the city of London with nodes and edge length representing the distances between the nodes \cite{karduni2016protocol}. For the experiments we took a sub-network of connected nodes with size 200 to mimic a possible arrangement of a skyway network. We then synthesize 2000 requests with random source and destination nodes. We also generate the payloads randomly with a maximum size of 10 packages and maximum weight of 5 kg for each package. We assume the drone takes 60 minutes to fully charge. We also assume that the drone speed is 65 km per hour. These variables are used to compute the energy consumption of the drone. The $charging time$ is the time a drone takes to charge to 100\% from its current state. The $waiting time$ is computed by summing the maximum charging times of concurrent charging drones while the others are waiting. Table \ref{tab:my-table} summarizes the experimental variables. 

\begin{table}[btp]
\small
\caption{Experiment Variables}
\label{tab:my-table}
\resizebox{\linewidth}{!}{%
\begin{tabular}{l|l}
\hline
Variable                                      & Value     \\ \hline
No. of nodes in the network subset            & 600 nodes \\
No. of nodes in the Largest Connected network & 200 node  \\
No. of generated Requests                     & 2000      \\
Max No. of packages in a Request              & 10        \\
Max weight of a package                       & 5 kg  \\
Time for a drone to charge from 0\% to 100\%  & 60 minutes\\
Battery consumption rate with 5kg payload      & 1\%/10 km \\
Speed of the drone                            & 65 km/hour\\

\hline    
\end{tabular}%
}
\end{table}

\begin{figure}[btp]
\centering
\includegraphics[width=0.9\linewidth]{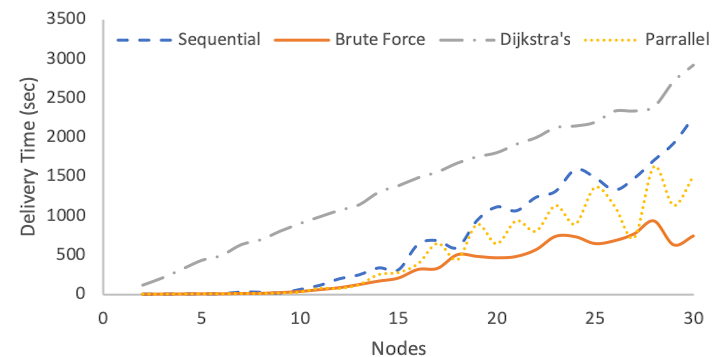}
\setlength{\belowcaptionskip}{-10pt}
\caption{Average delivery times for the proposed compositions, Dijkstra's algorithm, and Brute Force Approach}
\label{deliverytimeall}
\end{figure}

\begin{figure}[btp]
\centering
\includegraphics[width=0.9\linewidth]{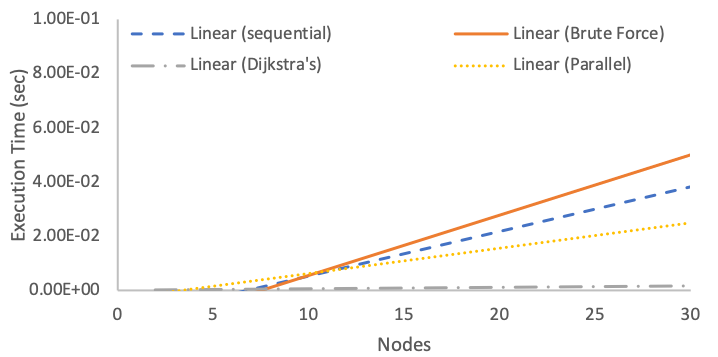}
\caption{Average execution times for the proposed compositions, Dijkstra's algorithm, and Brute Force Approach}
\label{executionall}
\end{figure}

\begin{figure}[btp]
\centering
\includegraphics[width=0.9\linewidth]{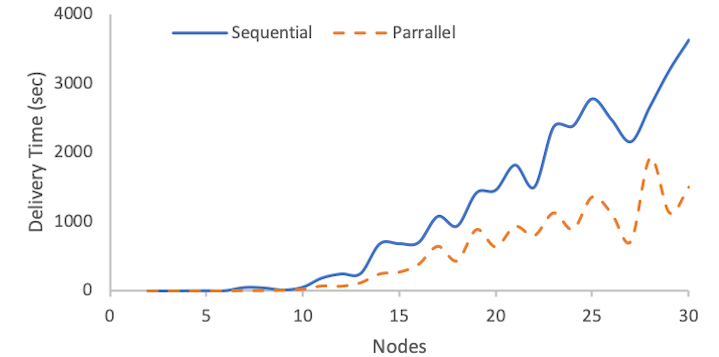}
\setlength{\belowcaptionskip}{-10pt}
\caption{Average delivery times for sequential, and parallel composition approaches}
\label{parallelvssequential}
\end{figure}

\begin{figure}[btp]
\centering
\includegraphics[width=0.9\linewidth]{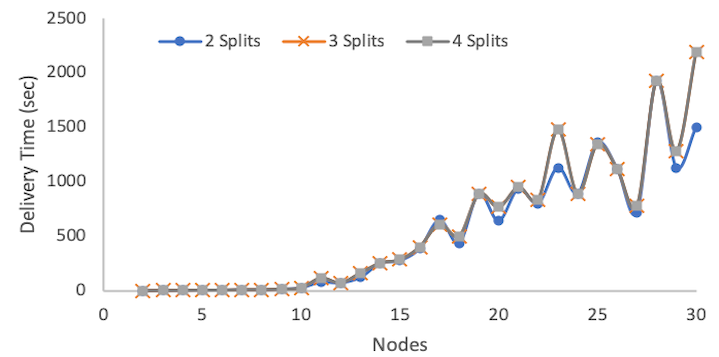}
\caption{Average delivery time of parallel composition approach varying maximum number of splits}
\label{splits}

\end{figure}

\begin{figure}[btp]
\centering
\includegraphics[width=0.9\linewidth]{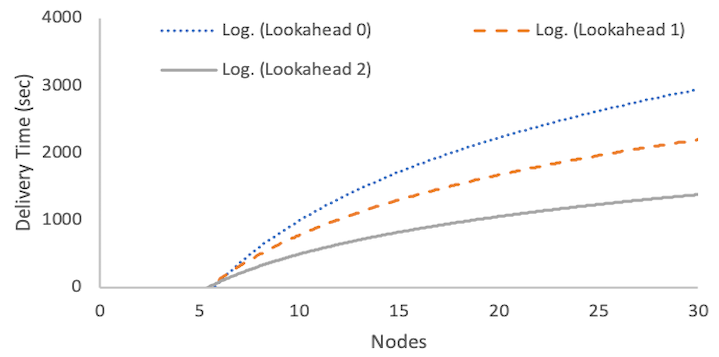}
\setlength{\abovecaptionskip}{-3pt}
\caption{Logarithmic trend line for delivery time of sequential composition approach with various lookaheads}
\label{radius}
\end{figure}

\begin{figure}[btp]
\centering
\includegraphics[width=0.9\linewidth]{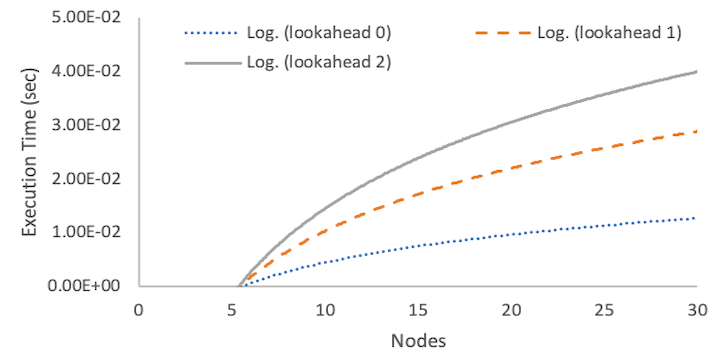}
\setlength{\abovecaptionskip}{-3pt}
\caption{Logarithmic trend line for execution time of sequential composition approach with various lookaheads}
\label{execution}
\end{figure}

\begin{figure}[btp]
\centering
\includegraphics[width=0.9\linewidth]{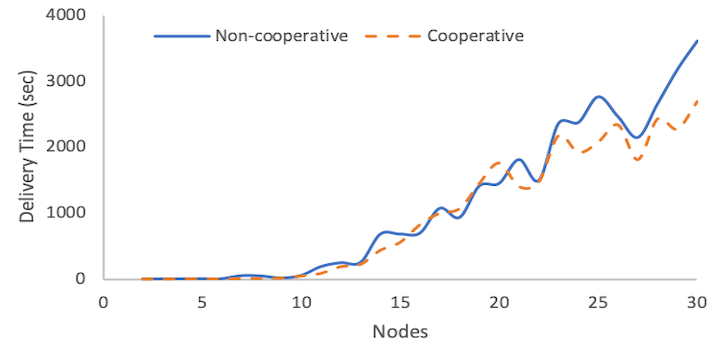}
\setlength{\abovecaptionskip}{-2pt}
\caption{Cooperation effect on Sequential Composition}
\label{sequentialaltruism}
\end{figure}

In the first experiment, we compare the average delivery times of the proposed composition algorithms against Dijkstra's algorithm \cite{dijkstra1959note} and the Brute Force approach. We consider the Brute Force approach as the baseline. For the Brute Force approach, we retrieve all the possible paths between the node and the destination in a request. Then, we run the SDaaS composition algorithm on every path and retrieve the path with the shortest delivery time. In the case of Dijkstra's algorithm, an edge represents the delivery cost, i.e. travel time + charging time + waiting time. We group the requests by the number of nodes between the source and the destination using the shortest path. We then compute the mean of each group. The x-axis in figures \ref{deliverytimeall} - \ref{sequentialaltruism} represent the number of nodes. Fig. \ref{deliverytimeall} shows the delivery times of the proposed algorithms with lookahead $l=2$, the Dijkstra's algorithm, and the Brute Force approach. As shown in the graph, the proposed algorithm performs better than the Dijkstra in terms of delivery time. Our proposed algorithm reduces the charging and waiting times significantly as it only charges the drones when they can't travel further. The Brute Force as a baseline performs the best but it comes with an exponential execution time cost as shown in the linear trend lines in Fig. \ref{executionall}.

For the Second experiment we focus on the average delivery time between the static and the dynamic swarm with $l=1$ and $max\_splits=2$. Fig.\ref{parallelvssequential} shows the trend for both sequential and parallel service compositions.  As shown in Fig.\ref{parallelvssequential}, the parallel composition reduced the delivery time significantly. This result supports our claim that parallel compositions maximize the utilization of the resources in the network. The maximized utilization results in a reduced charging and waiting times and an overall reduced delivery time. As the payloads of the requests are randomly generated, the graph shows a random behavior between nodes. The behavior is similar in both sequential and parallel compositions as the same set of requests are used.

In the third experiment, we focus on the parallel service compositions to measure the effect of maximum splits allowed at a node. If $max\_splits =2$, it means a swarm can disband to a maximum of 2 sub-swarms at a time. Fig. \ref{splits} shows the trend for a maximum of 2, 3, and 4 splits at a node. As we can see, increasing the maximum number of splits result in equal delivery time or a longer delivery time. This is because increasing the splits may result in increasing the dispersion of sub-swarms in the network. Our algorithm looks into the locally best neighboring node which could be of a further distance from the destination node. This behavior could be controlled if we have the actual GIS location of every node. Then, we can add a constraint to consider the actual distance between the neighboring node and the destination when selecting the best neighbor.

For the fourth experiment, we study the effect of increasing the lookaheads of neighboring nodes. The lookahead refers to the level of connection to the current node. A lookahead of 0 means all the neighbors considered are directly connected to the current node. Fig. \ref{radius} shows the general logarithmic trend line for the delivery time varying the lookaheads. As shown in Fig. \ref{radius} , as the lookaheads increases the delivery time decreases. This is because, the chance of selecting a better neighbor, i.e. least travel and charging times , increases.  However, including more neighbors comes with a computation cost. As the number of lookaheads increases the execution time also increases. Fig. \ref{execution}
 shows the logarithmic trend line for the execution time with various lookaheads.\looseness=-1
 
The final experiment measures the effect of \emph{cooperation} on the service composition. In this experiment, if the number of drones at a node are more than the number of charging pads, the drones only charge to what takes them to the next node instead of charging to 100\%. Fig. \ref{sequentialaltruism} shows the cooperative behavior in the sequential composition as it is more significant in this type since drones do not disperse in the network. The chances of having charging pads less than the number of drones in the full swarm is higher in a sequential composition. As shown in the figure, the delivery time is improved with the cooperative behavior especially when the number of nodes increases, when there are higher chances of recharges required.

\vspace{-0.2cm}

\section{Conclusion}
\label{conclusion}
\vspace{-0.2cm}
We propose a Swarm-based Drone-as-a-Service (SDaaS) composition framework for delivery services. We identified two types of service compositions namely: Sequential and Parallel. Two approaches are proposed for all types of compositions. The two approaches take all the constraints surrounding SDaaS into consideration. We then adopt Cooperation to enhance the delivery time by reducing the charging and waiting times. Experimental results show that the parallel composition outperforms the sequential composition. The efficiency of the proposed approach is proven against Dijkstra's and Brute Force approaches. It also shows that adopting a cooperative behavior improves the delivery time. The results also show the effect of varying the number of maximum splits allowed and the lookaheads of considered neighboring nodes from the current node. In the future work, we will consider extrinsic parameters such as weather conditions and extend the work to deal with SDaaS failures.






%

\def\IEEEbibitemsep{-1.2pt plus -15pt}

\vspace{-0.1cm}

\bibliographystyle{ieeetr}

\bibliography{swarmBased}


\cleardoublepage

\end{document}